\documentclass{ws-procs9x6}
\usepackage{graphicx}
\usepackage{amssymb}

\begin{document}

\title{QCD String formation and the Casimir Energy}

\author{K. Jimmy Juge}

\address{Institute for Theoretical Physics, \\
         University of Bern, \\
         Sidlerstrasse 5, \\
         CH-3012 Bern, Switzerland\\ 
         E-mail: juge@itp.unibe.ch}

\author{J. Kuti\footnote{\uppercase{S}peaker at the conference.}}

\address{Department of Physics, \\
         University of California at San Diego, \\
         La Jolla, California 92093-0319\\
         E-mail: jkuti@ucsd.edu}

\author{C. Morningstar}

\address{Department of Physics, \\
         Carnegie Mellon University, \\
         Pittsburgh, PA 15213, USA\\
         E-mail: cmorning@andrew.cmu.edu}


\maketitle

\abstracts{
Three distinct scales are identified in the
excitation spectrum of the gluon field around a static 
quark-antiquark pair as the color source separation R is varied. 
The spectrum, with string-like excitations on
the largest length scales of 2--3 fm, provides clues in its rich fine 
structure for developing an effective bosonic string description.
New results are reported from the three--dimensional Z(2) and SU(2) gauge 
models, providing further insight into the mechanism of bosonic string 
formation. The precocious onset of 
string--like behavior in the Casimir energy of
the static quark-antiquark ground state
is observed below R=1 fm where 
most of the string eigenmodes do not exist and the few stable 
excitations above the ground state are displaced.
We find no firm theoretical foundation for
the widely held view of discovering string formation 
from high precision ground state properties below the 1 fm scale.
}

\section{QCD String Spectrum and the Casimir Energy}

Last year, we presented a
new analysis of the fine structure in the QCD string spectrum
at the Lattice 2002 conference. Shortly afterwards, two papers were
submitted using complementary methods for
finding definitive signals of
bosonic string formation from the rich excitation spectrum\cite{JKM}
and the ground state Casimir energy.\cite{LW1} 

\vskip 0.1in
\noindent{\it QCD String Spectrum}
\vskip 0.05in

Three exact quantum numbers which are based on the symmetries of the problem
determine the classification scheme of the gluon excitation spectrum
in the presence of a static $q\bar q$ pair.\cite{JKM}
\begin{figure}[ht]
\centerline{
\includegraphics[width=3.0in]{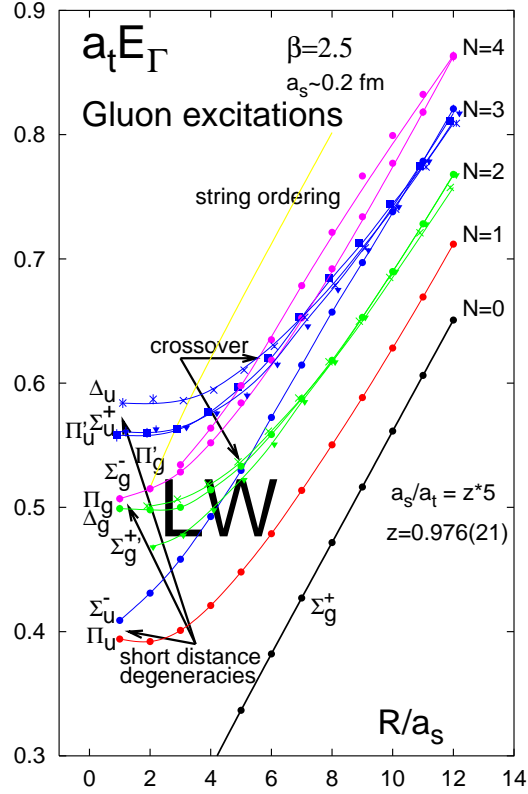}}
\caption{ Short distance degeneracies which are not string-like and 
their crossover
towards the QCD string spectrum are shown from Ref. 1 where further
details are explained. 
The color coded solid curves with simulation points, which identify
energy levels degenerate in the asymptotic string limit, are only shown for 
visualization and do not represent fits to the data.
The yellow line without data points marks a lower bound for the onset 
of mixing effects with glueball states which requires careful interpretation.
The symbol LW indicates the ${\rm R}$ range
of high precision Casimir energy calculations from Ref. 2. }

\label{fig:fig1}
\end{figure}
We adopt the standard notation from the physics of diatomic molecules
and use $\Lambda$ to denote the magnitude of the eigenvalue of the projection
${\bf J}_g\!\cdot\hat{\bf R}$ of the total angular momentum ${\bf J}_g$
of the gluon field onto the molecular axis with unit vector
$\hat{\bf R}$. The capital Greek
letters $\Sigma, \Pi, \Delta, \Phi, \dots$ are used to indicate states
with $\Lambda=0,1,2,3,\dots$, respectively.  The combined operations of
charge conjugation and spatial inversion about the midpoint between the
quark and the antiquark is also a symmetry and its eigenvalue is denoted by
$\eta_{CP}$.  States with $\eta_{CP}=1 (-1)$ are denoted
by the subscripts $g(u)$.  There is an additional label for the
$\Sigma$ states; $\Sigma$ states which
are even (odd) under a reflection in a plane containing the molecular
axis are denoted by a superscript $+$ $(-)$.  Hence, the low-lying
levels are labeled $\Sigma_g^+$, $\Sigma_g^-$, $\Sigma_u^+$, $\Sigma_u^-$,
$\Pi_g$, $\Pi_u$, $\Delta_g$, $\Delta_u$, and so on, $\Sigma_g^+$ 
designating the ground state.  For convenience,
we use $\Gamma$ to denote these labels in general.
For better resolution of the fine structure in the spectrum,
the gluon excitation energies ${\rm E_\Gamma(R) }$ were extracted from 
Monte Carlo estimates of generalized large Wilson loops on lattices with
small ${\rm a_t/a_s}$ aspect ratios and improved action.
Restricted to the R=0.2--3 fm range of a selected simulation,
the energy spectrum is shown in Fig.~\ref{fig:fig1} for 10 excited states.

On the shortest length scale, the excitations are consistent 
with short
distance physics without string--like level ordering in the spectrum.
A crossover region below 2 fm is identified 
with a dramatic rearrangement of the short distance level ordering.
On the largest length scale of 2--3 fm, the spectrum exhibits
string-like excitations with asymptotic ${\rm \pi/R}$ string gaps
which are split and slightly distorted by a fine structure. 
It is remarkable that
the torelon spectrum of a closed string, with one unit
of winding number around a compactified direction,
exhibits a similar fine structure on the  2--3 fm scale, 
as reported for the first time at Lattice 2003.\cite{JKMMP} 
This finding eliminates the boundary effects of fixed color charges as
the main source of the fine structure in the distorted spectrum.

\vskip 0.1in
\noindent{\it Casimir Energy}
\vskip 0.05in

In a complementary study,\cite{LW1} a string--like Casimir energy 
and the related effective conformal charge,
${\rm C_{eff}(R) = -12R^3F'(R)/(\pi(D-2))}$,
were isolated where ${\rm F(R)}$ is the force
between the static color sources and D is the space-time 
dimension of the gauge theory with bosonic string formation. 
With  unparalleled accuracy, ${\rm C_{eff}(R)}$ 
was determined 
for the gauge group SU(3) in D=3,4 dimensions, 
in the range ${\rm 0.2~fm < R < 1.0~fm}$, below the
crossover region of the string spectrum. 
A sudden change with increasing R, well below 1 fm, was observed 
in ${\rm C_{eff}(R)}$, breaking away from the 
the short distance running Coulomb law towards the string-like  
${\rm C_{eff}(R)\approx 1}$ behavior.
This was interpreted as a signal for
early bosonic string formation. 
The results are surprising because
the scale $\rm R$ is not large compared with the expected width of the
confining flux, and more quantitatively, the string-like 
Casimir energy behavior is
observed in the ${\rm R}$ range where the spectrum exhibits 
complex non-string level ordering, as shown in Fig.~\ref{fig:fig1}. 
We will try to develop now a better understanding of 
the seemingly paradoxical situation.

\section {Bosonic String Formation in the Z(2) Gauge Model}

The three-dimensional Z(2) lattice gauge model represents considerable
simplification in comparison with four-dimensional lattice QCD.
The SU(3) group elements on links are replaced by Z(2) variables, 
and the reduction to three dimensions implies 
a nontrivial continuum limit with a finite fixed point gauge coupling.

\vskip 0.1in
\noindent{\it Dual $\phi^4$ Field Theory Representation}
\vskip 0.05in

The main features are easily seen from the dual transformation 
of the Z(2) gauge model to Ising variables which can be replaced by the real
scalar field of ${\rm \phi^4 }$ field theory in the critical region.\footnote{
References to earlier work on 
the three-dimensional Z(2) gauge model can be found
in a recent paper on the finite temperature properties of 
the Z(2) string.\cite{CHP}}
The continuum model exhibits confinement and bosonic string formation
in the broken phase of the Ising representation. In addition,
a nontrivial glueball spectrum is observed\cite{CHPZ} with finite masses 
when measured in units
of the string tension. The string tension of the confining
flux in the Z(2) gauge model becomes the interface energy in 
the dual Ising-$\phi^4$ representation, and the lowest 
mass ${\rm 0^{+}}$ glueball state of the gauge theory with mass m maps into 
the elementary scalar
of the dual lattice, with inverse
correlation length m in the critical region. Higher glueball
states are Bethe-Salpeter bound states of the elementary scalar.\cite{CHPZ}
%%These are the most important generic ingredients
%%in the study of confinement, string formation, and the 
%%related string Casimir effect. 
The  dual $\phi^4$ Lagrangian of the Z(2) 
gauge model, in rigorous theoretical setting, is in analogy
with the dual Landau-Ginzburg superconductivity model which attempts
to describe the unknown microscopic quark confinement mechanism of QCD.
\footnote{ A recent review of quark confinement
and dual superconductivity
is given in Ref.~\refcite{Baker} with discussion of 
earlier work and references. }

The Ising-${\rm \phi^4}$ field theory model is particularly intriguing 
from the microscopic 
string theory viewpoint, if we recall Polyakov's work on the 
connection with the theory of random surfaces. Using
loop equations of closed Wilson loops near the continuum limit, he conjectured 
the equivalence of the three-dimensional Z(2)
lattice gauge theory
to a fermionic string theory.\cite{Polyakov}

\vskip 0.1in
\noindent{\it Renormalization Scheme}
\vskip 0.05in

In three euclidean dimensions, the Z(2) model is described in the
critical region (continuum limit) by a real order parameter
field $\phi$ with the Lagrangian
\begin{equation}
{\mathcal L} = {\rm -\frac{1}{2}(\partial_\mu\phi_0)^2 -
\frac{g_0}{4!}(\phi_0^2 - \frac{3m_0^2}{g_0})^2 } ~.
\end{equation}
The most frequently used renormalization
scheme requires in the broken phase that the 
tadpole diagrams completely cancel without
coupling constant renormalization ($g=g_0$) and with
wave function renormalization $\phi_0=\sqrt{Z}\phi$.
In the following, with lattice cutoff, we define a
scheme with finite
coupling constant renormalization keeping 
the renormalized mass of the elementary scalar exactly at
the pole of its propagator.
Since the wave function renormalization is finite to every order,
for convenience we choose Z=1 in 1-loop calculations.
With ${\rm g_0 = g +\delta g }$,
${\rm v^2_0\equiv 3m^2_0/g_0=v^2+\delta v^2}$, the
renormalized Lagrangian for elementary excitations $\eta$ around 
the vacuum expectation value $\phi=v$
is the starting point of the renormalized loop expansion with 
two counterterms to one-loop order,
\begin{equation}
 \delta {\rm v^2} = {\rm lim_{V\rightarrow \infty}\;\; \frac{3}{2}g \cdot
\frac{1}{V}} \sum_{\vec{\rm k}}\frac{1}{E^0_{\vec{\rm k}}}~, ~~~~
\delta{\rm g = \frac{g^2}{m}\;\frac{ln3}{32\pi}}~.
\label{eq:counter}
\end{equation}
The infinite spatial volume limit is taken in the sum over
the spectrum of inverse lattice energies
of free massive excitations $E^0_{\vec{\rm k}}$ 
with periodic boundary conditions.
The coupling constant counterterm ${\rm\delta g}$ satisfies the
renormalization condition on the physical propagator pole
to one-loop order.

In the presence of a pair of static sources, 
represented by a Wilson loop in the Z(2) gauge model, 
the renormalization
procedure is unchanged in the 
Ising-${\rm \phi^4 }$ field theory
description. The only change is in the
lattice Lagrangian where the sign of the nearest neighbor interaction
term is flipped on links which puncture
the surface of the Wilson loop on the dual Z(2) gauge lattice. 
This flip represents a disorder
line, or seam, between the two static sources on the spatial lattice. 
The end points
of the seam are fixed but otherwise it is deformable by a
``gauge transformation" of variables without changing the partition 
function. This invariance is inherited from the gauge invariant 
representation of the Wilson surface in the Z(2) gauge model.

\vskip 0.1in
\noindent{\it Numerical Implementation of the Loop Expansion}
\vskip 0.05in
The dual transformation of the Z(2) model to Ising
variables facilitates very efficient simulations with multispin coding.
The loop expansion provides theoretical insight into
Monte Carlo simulations of the excitation spectrum
using high statistics multispin Ising codes complemented by
${\rm \phi^4 }$ field theory codes.
Since the fixed point value ${\rm u^*}$ of the dimensionless coupling constant 
${\rm u=g/m}$ is not small, the simulations provide an important
cross-check on the convergence of the loop expansion which itself has to
be implemented in a numerical procedure.
The renormalized loop expansion in the presence of static
sources requires the following three steps.

\vskip 0.1in
\noindent {\bf (i)} First, for a given physical mass m and renormalized 
coupling g, the time independent renormalized classical field equation,
\begin{equation}
{\rm
  -\frac{\partial^2\phi_s(x,y)}{\partial x^2} -
   \frac{\partial^2\phi_s(x,y)}{\partial y^2} -\frac{1}{2}m^2\phi_s(x,y) + 
    \frac{g}{6}\phi^3_s(x,y) = 0 ~,
}
\label{eq:field_eq}
\end{equation} 
of the static soliton $\phi_s$ is solved on the lattice 
in the $\vec{\rm r}={\rm (x,y)}$ plane
with flipped nearest neighbor interaction
links along the seam between static sources.
In the Z(2) gauge model, the two sources can be interpreted as 
opposite sign charges with an electric flux connecting them.
In the ${\phi^4}$ representation we refer to the
$\phi_s$ classical solution as a static soliton, rather than the earthy
flux--tube term.
In the numerical procedure,
a generalized Newton type nonlinear iterative scheme was implemented
to obtain $\phi_s$ to double precision accuracy.

\vskip 0.1in
\noindent {\bf (ii)} Second, the fluctuation spectrum around 
the static soliton ${\rm \phi_s(x,y) }$
is determined by splitting the field into the classical solution plus
fluctuations, 
${\rm \phi(x,y,t) = \phi_s(x,y)}$ + ${\rm \eta(x,y,t)}$, with the 
eigenmodes of the fluctuation field
${\rm \eta(x,y,t)=\sum_n [ a_n(t)\psi_n(x,y) + a^\dag_n(t)\psi^*_n(x,y) ]}$ 
satisfying the eigenvalue equation 
\begin{equation}
{\rm
  -\frac{\partial^2\psi_n}{\partial x^2} -\frac{\partial^2\psi_n}{\partial y^2}
  + U''(\phi_s)\psi_n = E^2_n\cdot\psi_n ~.
}
\label{eq:fluct}
\end{equation} 
The time dependence of the fluctuation field $\eta$ is given in 
interaction picture by  
${\rm a_n(t) = a_n(0)e^{-iE_nt}}$ 
where the Hamiltonian is split into a
quadratic part and an interaction part of the $\psi_n$ eigenmodes.
The second derivative of the
$
{\rm U(\phi) = \frac{g}{4!}(\phi^2 - \frac{3m^2}{g})^2}
$
renormalized field potential energy
is taken with respect to $\phi$ in Eq.~(\ref{eq:fluct}), 
with $\phi=\phi_s$ substituted subsequently.
Two parity quantum numbers ${\rm P_x,~P_y}$ split
the eigenmodes into four separate symmetry classes.
With the two sources located at
${\rm (x,y)=(R/2,0)}$ and ${\rm (x,y)=(-R/2,0)}$, 
the quantum number ${\rm P_x=\pm 1}$ corresponds to the reflection symmetry
${\rm x\rightarrow -x}$ of ${\rm \psi_n(x,y)}$ and
${\rm P_y=\pm 1}$ corresponds to the ${\rm y\rightarrow -y}$
reflection symmetry. 
The full spectrum of eigenvalues and eigenfunctions of Eq.~(\ref{eq:fluct})
are computed by an Arnoldi diagonalization procedure in the finite volume
of the lattice. Using the parity symmetries of the theory, diagonalization
of large lattices with sizes up to 200x200 in the (x,y) plane were performed.

\vskip 0.1in
\noindent {\bf (iii)} Third, the systematic renormalized loop expansion
with the $\phi_s$ static soliton background is developed 
by building the finite volume field propagator 
${\rm D_F(}\vec{\rm r}{\rm,t;}\vec{\rm r}{\rm\,',t')}$ in Minkowski time,
\begin{equation}
{\rm D_F(}\vec{\rm r}{\rm,t;}\vec{\rm r}{\rm\,',t')}) = 
\sum_n {\rm i\int \frac{dp_0}{2\pi} }
\frac{\psi_n(\vec{\rm r})\psi^*_n(\vec{\rm r}\,')}{
{\rm p^2_0-E_n^2 + i\epsilon} }{\rm e^{-ip_0(t-t')} } ~,
\label{eq:prop}
\end{equation}
from the static $\psi_n(\vec {\rm r})$ eigenmodes. An euclidean rotation is
performed on the propagator during the numerical evaluation of the loop
diagrams.
The counterterms ${\rm \delta v^2}$ and ${\rm \delta g}$ 
are used to remove loop divergences 
in the continuum limit and
to keep the exact pole location at the physical mass m.
Using the propagator of Eq.~(\ref{eq:prop}), 
the fluctuation correction to the static soliton profile $\phi_s$ 
was calculated to one-loop order, together with similar calculations of the 
ground state energy and excitation energies. In this work, we only report
numerical results on the fluctuation spectrum of Eq.~(\ref{eq:fluct}) and its
1-loop contribution to the ground state energy.

\vskip 0.1in
\noindent{\it String Excitations in the Loop Expansion}
\vskip 0.05in

For sufficiently large R, the discrete ${\rm P_y=-1}$ bound state 
spectrum of Eq.~(\ref{eq:fluct}) is expected to evolve into
the asymptotic 
${\rm E_N = \pi N/R}$ ${\rm (N=1,2,\ldots})$ Dirichlet string spectrum of
massless string excitations which originate from 
the translational mode of the well-known one-dimensional $\phi^4$ soliton
by the following simple consideration.

Consider first the spatial lattice in the finite (x,y) 
plane with a seam of flipped links winding around the compact x--direction 
with periodic boundary condition. 
The classical solution 
${\rm \phi_{tor}(y) = m\sqrt{3/g}\;tanh(m|y|/2)}$ of Eq.~(\ref{eq:field_eq}) 
defines the torelon which is independent of x and
winds around the compact x--direction with a seam positioned at y=0.
We use continuum notation for the torelon and its
excitations, but finite cutoff and volume effects are
included in the numerical work. For ${\rm x>0}$, the 
transverse profile of the torelon
is identical to that of the well-known one-dimensional soliton, and
for ${\rm x<0}$ a sign flip is involved because of the seam at y=0.
The torelon eigenmodes of Eq.(\ref{eq:fluct}) with ${\rm P_y=-1}$ have the
simple form 
\begin{equation}
{\rm \psi^{tor}_{{}_N}(x,y) = 
\sqrt{\frac{g}{2m^3}}\;\phi_{tor}'(y)\cdot\sqrt{1/R}\;
e^{ip_{{}_N}\cdot x}},
\label{eq:toron}
\end{equation}
with quantized momenta, ${\rm p_{{}_N}=2\pi N/R, ~N=\pm 1,\pm2,\ldots}$, 
running along x in
the compact interval R with periodic boundary condition. The energy 
spectrum is given by ${\rm E_N = 2\pi N/R}$, with positive N values.

\begin{figure}[h,t]
\centerline{
\includegraphics[height=4.5in,angle=270]{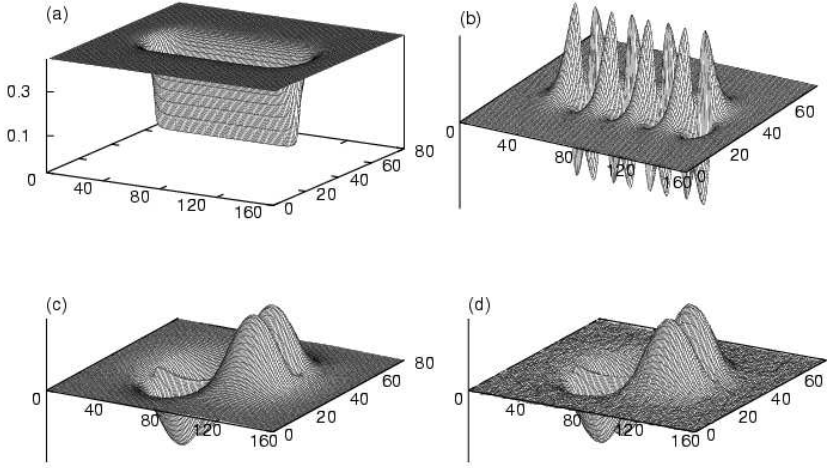}}
\caption{The static soliton solution $\phi_s$ of 
field equation (\ref{eq:field_eq}) is shown in (a)   
with the choice R=100 and renormalized
parameters given in the text; (b) shows the N=8 massless 
string excitation of the static soliton from the numerical
diagonalization of the eigenvaule equation (\ref{eq:fluct});
the second massive string 
excitation (K=2 breathing mode) is shown in (c) from the numerical solution
of the eigenvalue equation, and compared in (d) to
exact Monte Carlo simulation of the same state with remarkable agreement.
}
\label{fig:mf}
\end{figure}

The classical transverse profile 
${\rm \phi_{tor}(y)}$ of the torelon coincides, to a good approximation,
with that of the
static soliton $\phi_s$, if the separation
between the sources is large enough.
The static soliton profile $\phi_s$ does not interpolate from
-v at large negative y to +v for large positive y at fixed x because of the
flipped links along the seam. Rather, $\phi_s$ approaches 
v everywhere, far away from the seam line.
The eigenmodes of the fluctuation operator are restricted now
between the two sources and they are close to the form 
\begin{equation}
{\rm \psi^{static}_{{}_N}(x,y) = \sqrt{\frac{g}{2m^3}}\;
\phi_{tor}'(y)\cdot\sqrt{2/R}\;sin(\pi N/x)}, ~ 
-R/2\leqq x  \leqq R/2~,
\label{eq:psi_static}
\end{equation}
with N taking positive integer values.
The spectrum of these
standing waves is the same as that of a massless Dirichlet string 
oscillating in the (x,y) plane with fixed ends.
The excited eigenmodes
of the effective Schr\"odinger equation, 
like the one of Fig.~\ref{fig:mf}b for N=8,
are therefore in one-to-one correspondence with massless Dirichlet string 
oscillations.
The spectrum and the wave functions are expected 
to be somewhat distorted at finite R because of the distortions of the
effective Schr\"odinger potential around the sources. 

Representative examples of the numerical work are shown
in Fig.~\ref{fig:mf} where the static soliton solution ${\rm \phi_s(x,y) }$
on a 160x80 spatial lattice in the (x,y) plane
corresponds to source separation R=100 and physical mass
m=0.319 in the critical region (all dimensional quantities
are expressed in lattice spacing units).
For later comparisons, the lattice correlation length ${\rm m^{-1}}$ and 
the renormalized coupling g were chosen in the critical region to match 
one of our Monte Carlo simulations
with v=0.45 and string tension ${\rm \sigma=0.0101}$. 
The static soliton solution $\phi_s(\vec{\rm r})$ determines the attractive
potential energy of the effective Schr\"odinger eigenvalue problem
in Eq.~(\ref{eq:fluct}) which has a discrete bound state spectrum and
a nearly continuous dense spectrum above the glueball threshold m 
representing scattering states on the static soliton  
in the infinite lattice volume limit. 

%\pagebreak
%\vskip 0.1in
%\noindent{\it Massive String Excitations}
%\vskip 0.05in

The one--dimensional soliton, with classical mass 
${\rm 2m^3/g}$,  has a massive
intrinsic excitation, or breathing mode, whose
excitation energy is ${\rm (\sqrt{3}/2)m}$. 
In the large R limit, the intrinsic excitations of $\phi_s$ with
${\rm P_y=+1}$
become massive breathing
modes of the Dirichlet string.  The asymptotic
spectrum of a massive string, given by
${\rm E_K = \sqrt{3m^2/4 +\pi^2 K^2/R^2},~
K=1,2,\ldots}$, is associated with eigenmodes
like the K=2 wave function
of Fig.~\ref{fig:mf}c. The corresponding standing wave solutions,
\begin{equation}
{\rm \psi_{{}_K}(x,y) = \sqrt{3m/4}\;
sech(my/2)\;tanh(m|y|/2)\cdot \sqrt{2/R}\;sin(\pi K/x)} ~,
\label{eq:massive_wf}
\end{equation}
originate from the massive excitations of the torelon
with restriction to standing waves in the ${\rm -R/2\leqq x  \leqq R/2}$
interval.

%\vskip 0.1in
\section{Probing the String Theory Limit}
%\vskip 0.05in

We describe the $\eta(\vec{ \rm r},{\rm t})$ fluctuations 
around the static soliton $\phi_s$ by the sum of three fields, 
$\eta(\vec{ \rm r},{\rm t})  =  \xi(\vec{r},t) + 
    \chi(\vec{r},t) + \varphi(\vec{r},t)$, with
\begin{eqnarray*}
{\rm \xi(x,y,t)} & = & {\rm \sum_N [ a_N(t)\psi_N(x,y) + 
a^\dag_N(t)\psi^*_N(x,y) ]}~,  \\
 {\rm \chi(x,y,t)} & = & {\rm \sum_K [ a_K(t)\psi_K(x,y) + 
a^\dag_K(t)\psi^*_K(x,y) ]}~,  \\
 {\rm \varphi(x,y,t)} & = & {\rm  \sum_n [ a_n(t)\psi_n(x,y) + 
    a^\dag_n(t)\psi^*_n(x,y) ]}~, 
\end{eqnarray*}
where $\xi$ is restricted to bound states with
negative ${\rm P_y}$ parity
which are expected to evolve into massless string excitations for large R.
The field $\chi$ is restricted to ${\rm P_y=+1}$ parity bound states
which evolve into massive string excitations, and $\varphi$ 
is a sum over scattering states above the ${\rm 0^+ }$ glueball threshold m
in the continuum.

These fields are coupled in the interaction Lagrangian, and when 
the massive fields $\chi$ and
$\varphi$ are integrated out, we get a nonlocal 
Lagrangian in the ${\rm \xi(x,y,t)}$ field describing
massless string excitations in the large R limit.
As indicated by Eq.~(\ref{eq:psi_static}),  
the y--dependence in all the ${\rm P_y=-1}$ parity bound state
wave functions is approximately factored out in the large R limit.
Hence, the ${\rm \xi(x,y,t)}$
field can be replaced on large length scales by the field 
${\rm f(x,t)}$
which becomes the geometric string variable of low energy excitations 
measuring the displacements of the flux center--line in the y--direction
as a function of x and t. 

\vskip 0.1in
\noindent{\it Effective String Action}
\vskip 0.05in

The nonrenormalizable effective action of the ${\rm f(x,t)}$
field, with the massive
fields integrated out, is given in a derivative expansion by
\begin{eqnarray}
 {\rm S_{eff}(f)} &=& -\int {\rm dxdt
 \bigg[ \frac{1}{ 2 \pi \alpha '}\; \left ( 1 - \frac{1}{ 2} (\partial f)^2
- \frac{1}{ 8} ((\partial f)^2)^2 - 
\frac{1}{16} ((\partial f)^2)^3 + \cdots \right ) } \nonumber \\
&&{} + 
{\rm  const\cdot (\partial f)^2 \Box (\partial f)^2 + \cdots  \bigg] } \; ,
\label{eq:string_action}
\end{eqnarray}
where $\partial \sim m^{-1}$ is a long wavelength expansion.
The string tension $\sigma$ can be expressed 
as ${\rm \sigma = (2 \pi \alpha ')^{-1}}$,
and the notation ${\rm (\partial f)^2 = \partial_\mu f\partial^\mu f}$ 
is used in Eq.~(\ref{eq:string_action}).
Since  $f$ is related to massless Goldstone excitations,
originating from the restoration of translation invariance in torelon
quantization, 
only derivatives of f appear in ${\rm S_{eff} }$.
The first three terms in the derivative expansion 
come from the kinetic terms in the original $\phi^4$ field theory action.
They are independent of the details of the field potential except 
for the overall factor of ${\rm (2 \pi \alpha ')^{-1}}$.
The first line in Eq.~(\ref{eq:string_action}) 
agrees with the equivalent terms of the Nambu-Goto (NG) action, 
${\rm S_{NG} = {\rm (2 \pi \alpha ')^{-1}}\sqrt {1 - (\partial f)^2} }$,
when expanded in ${\rm (\partial f)^2}$.
The second line in Eq.~(\ref{eq:string_action})
has contributions from the geometric curvature, but new
contributions also appear whose geometric origin remains unclear.
This is where the effective string action begins to show deviations
from the NG string. 
To correct for end effects around the static sources, the effective action of 
Eq.~(\ref{eq:string_action}) has to be augmented by boundary operators
for the complete description of the Dirichlet string.

\vskip 0.1in
\noindent{\it Exact Excitation Spectrum from Simulations }
\vskip 0.05in

First, the physical mass, the vacuum expectation value v, and
the renormalized coupling g were determined in 
high precision Monte Carlo simulations of the bulk euclidean lattice action
in the Ising-${\rm \phi^4 }$ field theory representation
without the seam of flipped links. The renormalized physical parameters
were used as input to compare the loop expansion with 
simulations in the presence of static sources.
The excitation spectrum around the static sources was determined from
Monte Carlo estimates of correlation matrices which included an extended
set of optimized operators. These operators were built from eigenfunctions 
of the fluctuation operator on two-dimensional time slices of the lattice.

\begin{figure}
\centerline{
\includegraphics[width=4.5in]{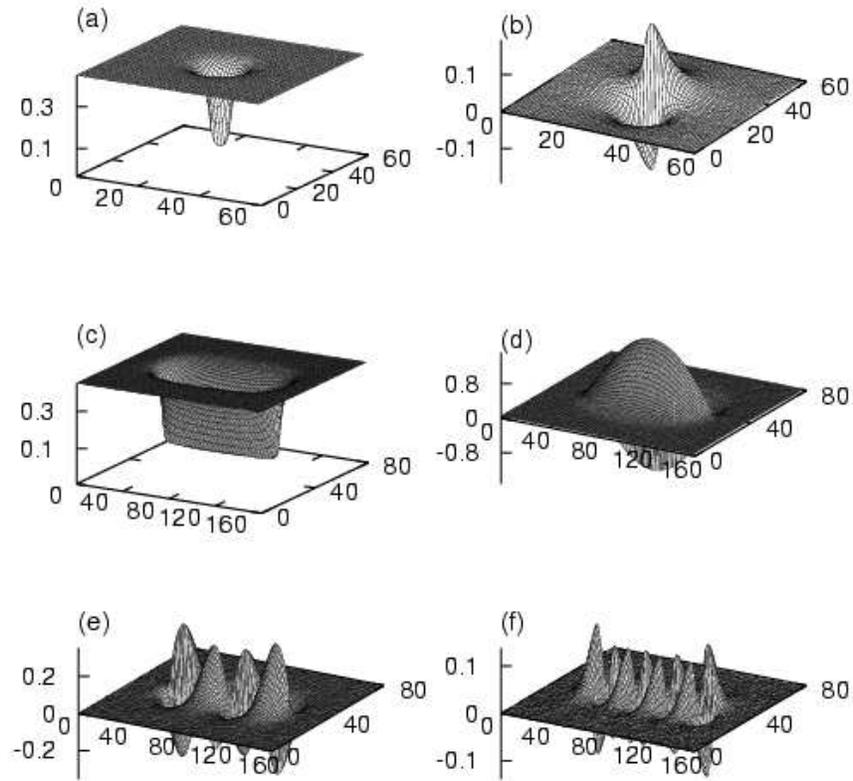}}
\vskip -1.5in
\caption{The soliton profile for R=10 is shown in (a), 
with the only bound state wave function below the glueball threshold 
depicted in (b).
At R=100, the stretched soliton configuration shown in (c) 
exhibits several string-like excitation with the N=1 wave function
shown in (d), N=4 in (e), and
N=8 in (f). The simulation results match the loop expansion of 
Fig.~\ref{fig:mf} with common renormalized parameters.
}
\label{fig:wf}
\end{figure}

Simulation results are shown in Fig.~\ref{fig:wf} for string
formation as R is stretched. The
shape of the static soliton profile ${\rm \phi_s(x,y)}$  
around the sources is depicted in the (x,y) plane of the two-dimensional
spatial lattice for R=10 and R=100.
With the scale set by the string tension,
the two R values correspond to 
0.3 fm and 3 fm, respectively. The renormalized
bulk physical parameters of the simulations were
given in the discussion of Fig.~\ref{fig:mf} as m=0.319, v=0.45,
and ${\rm \sigma=0.0101}$. 
At the smaller R value,
the static soliton with bag--like shape is not stretched, and there is 
only one bound state excitation below the glueball threshold.
At R=3 fm, the stretched static soliton supports many string
excitations, with the N=1,4,8 wave functions displayed. The 
Monte Carlo simulations 
are in good agreement with results from the eigenmodes of the
fluctuation operator.  This is indicated by the good match of
Fig.~\ref{fig:mf}a versus Fig.~\ref{fig:wf}c and
Fig.~\ref{fig:mf}b versus Fig.~\ref{fig:wf}f.
\begin{figure}
\centerline{
\includegraphics[height=4.0in,angle=-90]{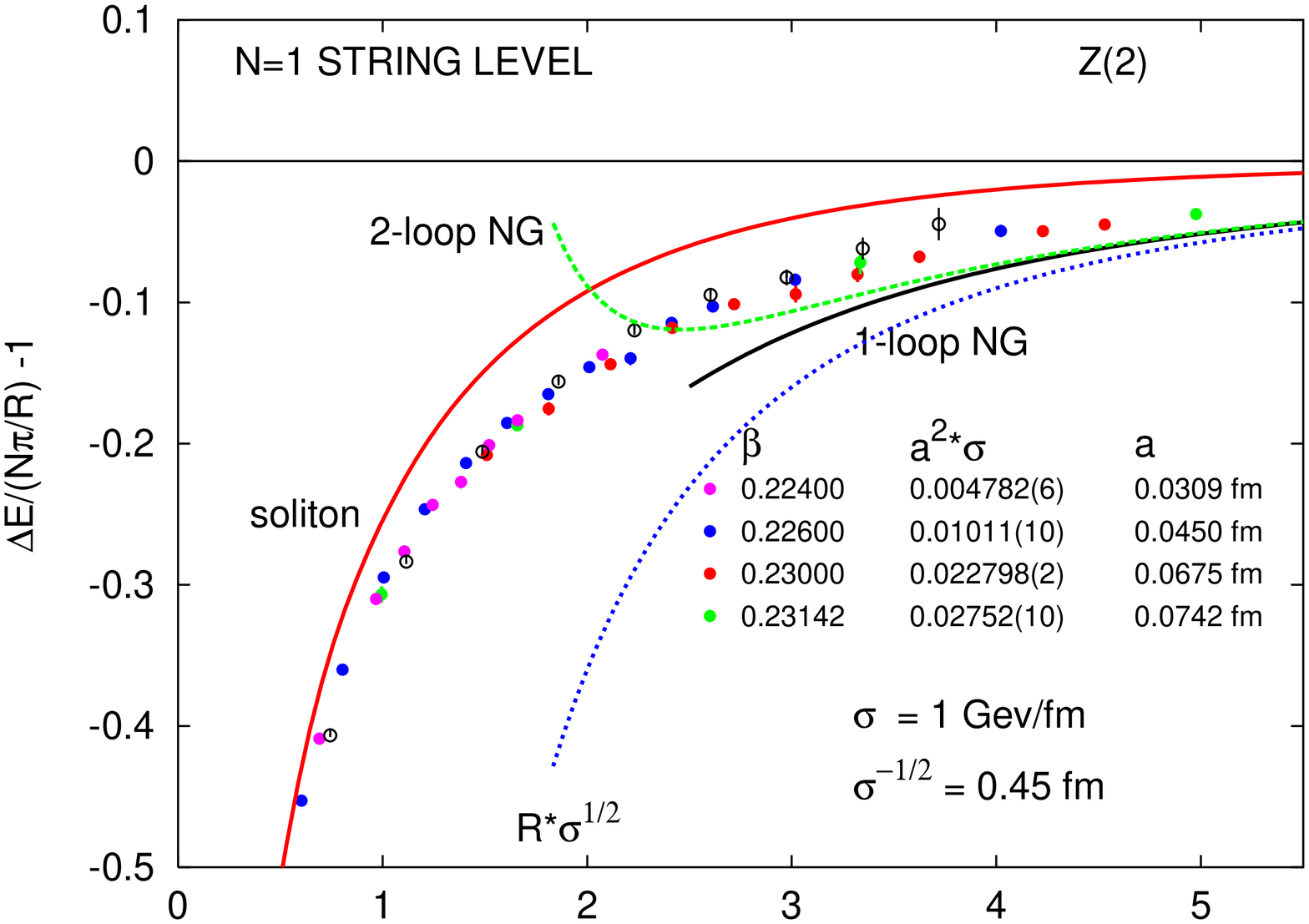}}
\caption{
The energy gap ${\rm \Delta E}$ above the ground state is plotted
as ${\rm \Delta E/(N\pi/R) - 1}$ to show percentage deviations   
from the asymptotic ${\rm N=1}$ string level.  Several Z(2) simulations
with cyan, blue, red, and green points
are combined with good scaling properties. The open circles 
represent D=3 SU(2) results after
readjusting the ratio of the string tension $\sigma$ to the glueball mass
in Z(2). The null line coresponds 
to the tree level  ${\rm \pi /N}$  NG string gaps.
The dashed blue and green lines are 1--loop and two--loop
NG approximations, repectively, and the black line is the full NG
prediction.
}
\label{fig:gaps1}
\end{figure}
\vskip 0.1in
\noindent{\it Matching the Excitation Spectrum to the String Action }
\vskip 0.05in

A large number of excitation spectra were obtained 
from highly accurate simulations as R was
varied in a wide range from 0.3 fm to approximately 10 fm.
A good test for string formation is provided by the 
behavior of the spectrum as a function of R.
The NG spectrum,
\begin{equation}
{\rm  E_N = \sigma R \left( 1 - \frac{D-2}{12\sigma R^2}\pi+ \
  \frac{2\pi N}{\sigma R^2} \right)^{\frac{1}{2}} } ~, 
\label{eq:Arvis}
\end{equation}
with fixed end boundary conditions in D dimensions, was
calculated in Ref.~\refcite{arvis}. 
N=0 corresponds to the string ground state and positive integer N values
label the excitations of the Dirichlet string.
Although there exists an inconsistency in the quantization of
angular momentum rotations around the ${\rm q\bar q}$-axis at finite
${\rm R}$ values unless ${\rm D=26}$, the problem asymptotically 
disappears in the
${\rm R \rightarrow\infty}$ limit.\cite{arvis}
This is expected from the earlier discussion on string formation
in the loop expansion. Indeed, derived in a consistent D=3 $\phi^4$ field
theory, the first few terms of the effective string
action match the coefficients of their NG counterparts as seen  
in Eq.~(\ref{eq:string_action}). If a string limit is reached  
for a large enough R range, the expansion of ${\rm E_N}$ into 
inverse powers of R from Eq.~(\ref{eq:string_action}) should agree
with the simulations at least to one nontrivial order. 
At small R values the expansion will break down since the
mathematical NG string will be the inconsistent
description of the bag-like soliton and its excitations in three dimensions.

The comparison of simulations to string theory is illustrated 
in Fig.~\ref{fig:gaps1} where
the exact N=1 excitation is plotted against
the numerical spectrum of the fluctuation operator 
and the predictions of the Nambu-Goto string model.
The numerical spectrum from Eq.~(\ref{eq:fluct}) 
is the renormalized tree level starting point of the loop expansion in
${\rm \phi^4}$ field theory setting without assuming string formation. 
It is close in shape
and details to the exact results for the entire
R range, as shown by the solid red line with the soliton tag
in Fig.~\ref{fig:gaps1}.
It is expected that higher loop corrections will bring the agreement
even closer. The simulations, however, deviate substantially from
the predictions of the loop expansion in the NG string model, 
particularly at smaller ${\rm R}$ values below 1 fm where the loop 
expansion suddenly begins to diverge. 
The string formation for large R is clearly seen in the spectrum 
of Fig.~\ref{fig:gaps1} and all the other spectra we obtained, but 
further work is needed
for quantitative matching of the coefficients in the effective string
action.

Our results differ from the findings of 
Ref.~\refcite{CHP} where simulations
of the finite temperature Z(2) string were reported in good
agreement with the expansion of the NG model to 1--loop order
below the R=1 fm scale.

\section{The Casimir Energy Puzzle}
The breakdown of the effective string description below 1 fm
and the related Casimir energy paradox are
illustrated with the calculation of the ground state
energy from the renormalized fluctuation operator around the static
$\phi_s$ soliton.
In the renormalized loop expansion, the early onset of the 
string-like ${\rm C_{eff}(R) \approx 1}$ 
behavior in the R range below 1 fm is not associated 
with massless string eigenmodes which are mostly missing for
${\rm R\lesssim 1}$ fm. The results are consistent with the
direct simulations of ${\rm C_{eff}(R)}$
and the spectrum in our Z(2) model.
The simulations of Refs.~\refcite{JKM,LW1} present the same 
puzzle in QCD. 
If there is any physics associated with this puzzle, it remains unresolved.

\vskip 0.1in
\noindent{\it Casimir Energy from the Fluctuation Operator}
\vskip 0.05in
The soliton ground state energy ${\rm E_s}$ in Eq.~(\ref{eq:casi1}), 
\begin{eqnarray}
{\rm  E_s } & = & {\rm E^{cl}_s + 
\frac{1}{2}\sum_{\alpha}E_\alpha  - 
\frac{1}{2} } \sum_{\vec k} {\rm E}_{\vec k}^0 \nonumber \\
& - & {\rm \frac{g}{12}\delta v^2 \int 
\left [ \phi_s - \frac{3m^2}{g} \right ]dxdy
+ \frac{\delta g}{4}\int \left [ \phi_s - \frac{3m^2}{g} \right ]^2dxdy ~, }
\label{eq:casi1}
\end{eqnarray}
includes the
renormalized classical energy ${\rm E^{cl}_s}$, the sum of zero point
energies summed over all eigenmodes $\alpha$ of the fluctuation 
operator of Eq.~(\ref{eq:fluct}), and a sum over  
all momenta ${\vec k}$ in the zero point
energy of the bulk vacuum without sources.
\begin{figure}[b]
\centerline{
\includegraphics[height=4.0in,angle=-90]{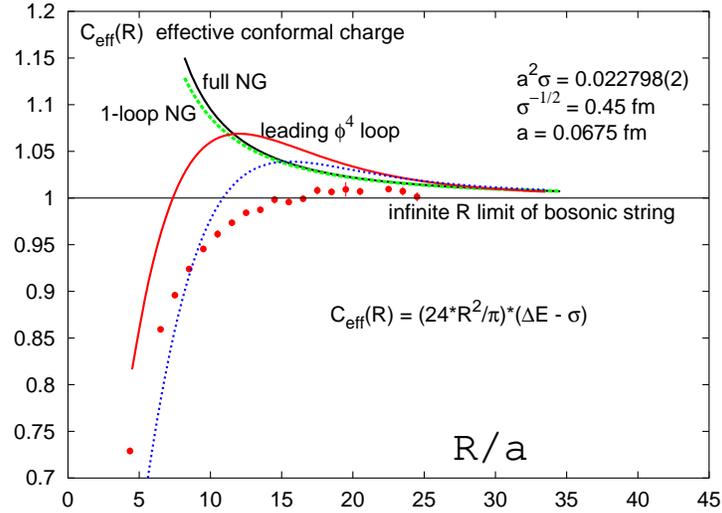}}
\caption{The red points are from high precision Ising-$\phi^4$
simulations. The solid black curve with NG label is the full NG
prediction, ${\rm C_{eff}(R) = 1 }$ is the asymptotic string 
result which corresponds to the tree-level NG prediction.
The dashed green line shows the 1-loop approximation which includes
the first correction to tree level from the ${\rm R^{-1} }$ expansion 
of Eq.~(\ref{eq:Arvis}). 
The solid red line is calculated from the numerical
evaluation of Eq.~(\ref{eq:casi1}). The dashed blue line is obtained from
the full red line by subtracting the ${\rm E^{cl}_s}$ contribution.
}
\label{fig:casi}
\end{figure}
\begin{figure}
\vskip -0.5in
\centerline{
\includegraphics[height=4.0in,angle=-90]{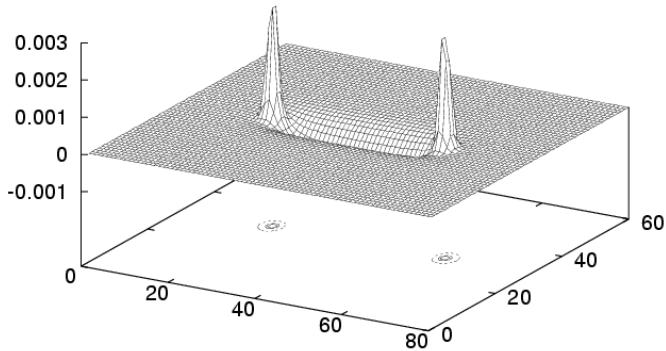}}
\caption{The ground state fluctuation energy density is plotted
for R=40. The two peaks represent large fluctuation contributions
around the static sources.  
}
\label{fig:casi_dens}
\end{figure}
The difference of the two eigenmode sums is still 
divergent in the continuum limit. The ${\rm \delta v^2}$ counterterm
removes this divergence and contributes a finite $\vec r$-dependent 
energy density to the ground state.
The ${\rm \delta g}$ counterterm also contributes a finite 
and $\vec r$-dependent ground state energy density.

\vskip 0.1in
\noindent{\it Exact Ground State Results}
\vskip 0.05in

The simple definition 
${\rm C_{eff}(R) = -24R^2(E_s'(R)-\sigma)/(\pi(D-2))}$
was used to isolate the effective Casimir energy term in the ground
state of the soliton. The first derivative ${\rm E_s'}$
was directly simulated on the lattice by a special method we developed.
The string tension $\sigma$ was determined in high precision
separate runs from the ground state of long torelons.
The simulations in Fig.~\ref{fig:casi} are also compared with
the predictions of the Nambu-Goto (NG) string model
and the predictions of Eq.~(\ref{eq:casi1}) from the numerical
evaluation of the fluctuation operator spectrum.

The agreement of ${\rm C_{eff}(R)}$ with exact simulation results, 
as determined from the fluctuation operator spectrum,
is quite good down to very small R values.
At smaller R values, full agreement can only be expected 
from higher loop corrections. 
It is interesting that the classical contribution of $\phi_s$ to 
${\rm C_{eff}(R)}$ is significant below 1 fm.
At large R, in the true string 
formation limit, the higher loop corrections should not contribute to 
the asymptotic 1/R Casimir term.

The dominant contribution to the ground state energy of the static 
soliton from the fluctuation operator is coming from 
the continuum spectrum but the ground state energy density remains
concentrated around the static soliton. This is also seen in
the Casimir energy density 
which is calculated from the first derivative of the energy density 
with respect to R, integrating to ${\rm C_{eff}(R)}$ as we explicitly checked.
For illustration,
the Casimir energy density is shown in Fig.~\ref{fig:casi_dens}.

Although the 1/R expansion of ${\rm C_{eff}(R)}$ from the NG
prediction of Eq.~(\ref{eq:Arvis}) is not divergent below 1 fm, 
it breaks away from the data in a rather dramatic fashion.

\section{Conclusions}

We established bosonic string formation in a large class of 
gauge theory models from a direct study of the excitation spectrum
at large separation of the static sources.
The spectrum, with string-like excitations on
the length scales of 2--3 fm and beyond, provides clues in its rich fine 
structure for developing an effective bosonic string description.
The matching of the string-like spectrum to an effective string
action remains a challenge.

Our results at small R differ from the findings of 
Ref.~\refcite{CHP} where simulations
of the finite temperature Z(2) string were reported in good
agreement with the expansion of the NG model to 1--loop order
below the R=1 fm scale. 
This agreement was interpreted as further
support at finite temperature for the precocious onset of bosonic
string formation in QCD below the 1 fm scale as 
reported in Ref.~\refcite{LW1}.

We find no firm theoretical foundation for
discovering string formation 
from high precision ground state properties below the 1 fm scale.
The explanation for string--like
finite temperature free energy
behavior below 1 fm also remains unclear. Further work is needed to 
understand the interpretation of the results 
from Refs.~\refcite{LW1,CHP} which present impressive 
high precision simulations.

\section*{Acknowledgments}
This work was supported by the DOE, Grant No. DE-FG03-97ER40546, the 
NSF under Award PHY-0099450, and the European
Community's Human Potential Programme under contract HPRN-CT-2000-00145,
Hadrons/Lattice QCD.

%%%%%%%%%%%%%%%%%%%%%%%%%%%%%%%%%%%%%%%%%%%%%%%%%%%%%%%%

\end{document}